\documentclass{aa}
\usepackage{graphics}

\def\Teff{${\rm T_{eff}~}$}
\def\logg{log~{\sl g~}}
\def\vt{$\rm v_{t}~$}
\def\kms{${\rm km \: s^{-1}}$}
\def\Vr{$\rm Vr_{helio}$}
\def\NLi{\rm N(Li) }

\begin{document}

\thesaurus{08.01.1; 08.12.1; 10.15.2 Berkeley~21}

\title{A Super Lithium Rich giant in the metal-poor open cluster Berkeley~21
\thanks{Based on observations collected at
the European Southern Observatory, La Silla, Chile.}}
\subtitle{}

\author{V. Hill\inst{1}, L. Pasquini\inst{1}}
\institute{ESO, Karl Schwarschild str. 2, D-85748 Garching b. M\"unchen\\
	   email: vhill@eso.org}

   \offprints{V. Hill}

   \date{}

   \maketitle

   \begin{abstract}

We present the analysis of FEROS commissioning spectra of 3 giants
in the metal poor cluster Be~21. One of the giants has an exceptionally high
Li content, comparable to the original Li in the solar system.
These objects are very rare (only a handful are
known), and this is the first Super Lithium Rich giant
(SLIR) discovered in an open cluster. The reasons for such a high Li abundance
are unknown: it could be the result of a short lived internal process,
or of accretion from external sources, the former being slightly
more likely.
From the spectra, the metal abundance is also derived for 3 giants,
giving a mean of $\overline{\rm [Fe/H]}=-0.54 \pm 0.2 \rm dex$, 
in good agreement with recent  photometric estimates, but 
substantially higher than estimates previously obtained.


   \end{abstract}

\section{Introduction}

Super Lithium Rich giants (SLIR) are an elusive class of objects,
serendipitously discovered.
The presence of a substantial amount of Li in `normal' giants 
is already a puzzle, because, according to current evolutionary theories,
very little Li should have remained in the stellar atmosphere
of these stars shortly after they leave the main sequence, and they 
are not expected to produce any substantial
amount before the Cameron-Fowler (\cite{cameron fowler 71}) mechanism becomes active,
i.e. before the star gets to the tip of the giant  branch and onto the
AGB.

In their survey for Lithium Rich giants, Brown et al. (\cite{brown etal
89}) found out of 644 giants, 8 Lithium Rich giants 
($\NLi \ge 1.2$~dex\footnote{$\NLi=\log {\rm (n_{Li}/n_{H})} + 12)$}) and only
2 with lithium abundances close to the current interstellar value
($\NLi \sim 3$~dex), while the compilation by de la Reza (\cite{de la
resa etal 96},\cite{de la resa etal 97}) quote 
$\sim$40 Lithium Rich giants.
With the exception of a recently discovered red giant belonging to the 
globular cluster M3 (Kraft et al. \cite{kraft etal 99}), all the other
giants with $\NLi \sim 3$~dex belong to the field, and what we report here 
is the first detection of such an object in an open cluster. 
This should help us
to precisely position the star in the colour-magnitude diagram, to
assess its evolutionary status, initial mass, age and metallicity.

Among possible theories of Li enhancement, pollution from an evolved
companion, engulfing of a lithium rich low mass (brown dwarf) companion,
self production and shell detachment (de la Reza et al. \cite{de la
resa etal 96},\cite{de la
resa etal 97})
have been proposed.

\section{Observations and Sample}

Four giants in the core of Be~21 were observed with FEROS,
the new fibre-fed high resolution echelle spectrograph that has been 
installed in 1998 on the 1.52m telescope in La Silla and has 
proven to be very stable and sensitive. The spectra were obtained 
during the commissioning period in November 1998 (Kaufer et al. 
\cite{kaufer etal 99}).  
Several exposures were obtained per star and the
data were reduced by using the FEROS on line data reduction system. 
Since FEROS also provides simultaneous sky spectra 
(from a dedicated fiber),
for these relative faint stars the sky spectra were used
to check for the presence of emission lines in the regions of
interest. No detectable  sky continuum was found, and therefore we
preferred not to subtract the sky spectra , in order not to
lower the S/N ratio of the spectra which is $\rm S/N \sim 20/pixel$ (a
quite remarkable result, considering that FEROS is fed by the 
ESO 1.52m telescope).
FEROS resolving power is of ${\rm R}=48000$ and FEROS is a rather
stable instrument, providing accurate radial velocities.

A brief logbook of these observations is given in
Table~\ref{T-logbook}, where 
the identification of the stars follows the numbering of Tosi et
al. (\cite{tosi etal 98}).

\begin{table}
\caption{Logbook of the observations} \label{T-logbook}
\begin{tabular}{lc@{}c@{}c@{}c@{}c}
\noalign{\smallskip}
\hline
\noalign{\smallskip}
 star &  $V$ &$(B-V)$&~$(V-I)$& Exp. time &\Vr\\
&&&&& (\kms)\\
\noalign{\smallskip}
\hline
\noalign{\smallskip}
T33  & 15.85 & 1.70 & 2.11 & 2$\times$1h  & 12.5\\ 	
T27 & 15.66 & 1.70 & 2.06 & 2$\times$1h  & 12.8\\	
T26 & 15.62 & 1.77 & 2.07 & ~2$\times$1h + 45mn& 12.6\\	
T9  & 15.00 & 1.95 & 2.29 & 2$\times$1h  & 11.5\\	
\noalign{\smallskip}
\hline
\end{tabular}
\end{table}

The radial velocities of the 4 giants are also given in
Table~\ref{T-logbook}.
The four radial velocities agree clearly very well
at a \Vr~ of 12.35$\pm 0.6$~\kms, 
which is compatible to the
intrinsic accuracy of FEROS without simultaneous calibration. 
We therefore conclude (at odds with
previous, less accurate measurements by Friel \& Janes (\cite{friel
janes 93}) that the four giants are likely Be~21 members and that the
radial velocity of the Be~21 cluster is 12.35~\kms.

The spectra were acquired in a short span one after the other
(at most one night difference) and we do not find noticeable radial velocity
variations among the different spectra.
However, the brightest giant (T9) shows a very broad an slightly
assymetric correlation function, which reflects the complex and broad
structure of its lines. We suspect that this star might be a binary,
and excluded it from our abundance analysis.

\begin{figure}[tbp]
\resizebox{\hsize}{!}{\rotatebox{-90}{\includegraphics{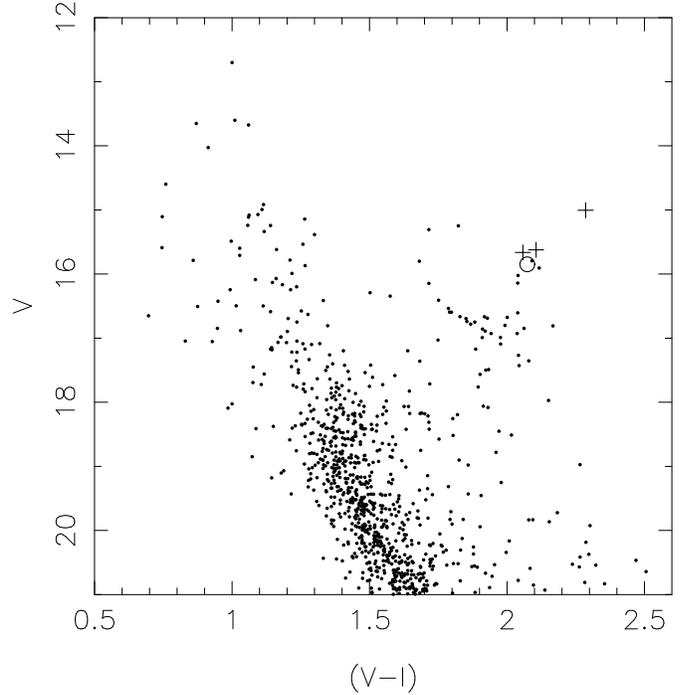}}}
\caption{Observed Colour-Magnitude diagram of the cluster Be~21 from
         Tosi et al. \cite{tosi etal 98}. The four stars studied here are plotted as
         crosses and circle respectively for the normal and the SLIR giant.}
         \label{F-hr}
   \end{figure}

The position of the four stars in the Colour-Magnitude diagram of Be~21
of Tosi et al. (\cite{tosi etal 98}) is shown in Figure~\ref{F-hr}, 
with the SLIR star marked as a circle. 
Clearly, the colours and magnitudes of the four giants
fit very well in the C-M diagram of the cluster and the three
fainter giants observed are about 1 magnitude above the
He-burning core clump. According to Tosi et al. (\cite{tosi etal 98}) and their
visual magnitudes, the absolute visual magnitude of the three stars is
of $\rm M_v \sim 0$.

\section{Analysis}

	\subsection{Stellar atmosphere models}

The model atmospheres that we used to derive the abundances  were interpolated
in a grid of models calculated with codes derived from the MARCS',
and especially
adapted to represent the atmosphere of low gravity stars~:
Plez's (\cite{plez etal 92} and \cite{plez 97}) grid of models,
with temperature and gravity ranges (3500$\leq$\Teff$\leq$4750~K and
 --0.5$\leq$\logg$\leq$1.0~dex) for metallicities --1.0, --0.6, --0.3, 0.,
0.3, 0.6~dex.

\subsection{Effective parameter determination} \label{effective parameters}

To determine the effective parameters of  the
stars the following procedure was adopted:
\begin{itemize}
\item{\em Temperature:}
the $(B-V)$ and $(V-I)$ colours of the star are used to determine the
effective temperature \Teff, which is then checked
``spectroscopically'', imposing  for the excitation equilibrium to be
achieved (lines with high and low
excitation potential should produce the same abundance for a given element
-here Fe~I).
\item{\em Surface gravity:}
\logg is derived from the  ionisation equilibrium: we require that the
iron abundance deduced from neutral and ionised species should be the same.
\item{\em Microturbulent velocity:}
\vt is derived imposing that the iron lines with small and large
equivalent width should give the same abundance.
\end{itemize}

To derive the input T$_{eff}$ we have started from the Tosi et al. values
of reddening: $E(B-V)=0.76$ (thus $E(V-I)=0.99$) and we have applied these
values to the stars. The effective temperature calibration of Alonso et al.
(\cite{alonso etal 96}) has been then applied; in fact, even if this
calibration has been mostly developed for dwarf stars,
models clearly show that in this range of temperature, 
it has a very little dependence on gravity, if at all (cf Bessell et
al. \cite{bessell etal 98}). 
The initial guess was found very close to the
spectroscopic results, as shown in Table~\ref{T-parameters},
confirming indirectly the quality of the of the Tosi et al. 
reddening analysis. 
In the same table,
we note , however, that the spectroscopic and photometric gravities
do not match very well, in that spectroscopic gravities (derived from
ionisation equilibrium) give
lower values than the photometric ones. This effect is well
known in intermediate metallicity stars and it is due to the
fact that the Fe lines suffer of NLTE effects (mainly
overionisation). In this case,
however, the effect is larger than expected, and this might be due to
the extremely small number of ionised lines that could be measured
(weak lines with $\rm W < 50$m\AA~ were impossible to measure reliably
because of the S/N of the spectra).
Since the stars belong  to a cluster, on the other hand we know that
the real gravities cannot be lower than the ones derived from the
photometry by more than a
factor 2. We have therefore in the following
the photometric gravities. This implies that
the Fe~II lines were not used to determine the mean [Fe/H] abundance.

Table \ref{T-parameters} summarizes the various determinations of 
 effective parameters for each star.

\begin{table}
\caption{Adopted effective parameters} \label{T-parameters}
\begin{tabular}{lc@{}c@{}cc@{}c@{}c}
\noalign{\smallskip}
\hline
\noalign{\smallskip}
Star 	& \multispan3 \hfill \Teff (K) \hfill & \multispan2
\hfill \logg (dex) \hfill& \vt \\
	& $(V-I)$&~$(B-V)$&~Spectr.&Phot.&~Spectr. &~(\kms)\\
\noalign{\smallskip}
\hline
\noalign{\smallskip}
T33	& 4800	& 4600	& 4600 	& 2.2	& $\leq$1. &2.0\\
T27	& 4800	& 4630	& 4600	& 2.2	& $\leq$1.	&2.0\\
T26	& 4630	& 4520	& 4500	& 2.0	& 1.5	&2.0\\
T9	& 4300	& 4380 	&  --	& 1.7	& -- 	& --\\
\noalign{\smallskip}
\hline
\noalign{\smallskip}
\end{tabular}
\end{table}

	\subsection{Metallicity} \label{atomic lines}

Metal abundances were derived by
measuring the  equivalent width of isolated lines by the Gaussian
approximation and deducing the abundances from these measurements.
We have also checked with the curve of growth method, finding
consistent results.  
In Table~\ref{T-abundances} the abundance of elements analysed
together with the number of 
lines and the scatter obtained are given.

The mean iron abundance  of the three stars is
 $\overline{\rm [Fe/H]}= -0.54$~dex. This metallicity is  larger
 than what had been found in previous studies, from photometric arguments
 or medium-resolution spectroscopy (Friel \& Janes \cite{friel janes 93}), but
in good agreement with the results obtained by Tosi et al (\cite{tosi etal 98}).
Being Be~21 one of the most metal poor clusters known (Friel \cite{friel 95}),
this result may have an impact on the models of Galactic chemical evolution.
We also note that a [Fe/H] of -0.9~dex as proposed by Friel \& Janes
(\cite{friel janes 93}) would require
adopting drastically lower effective temperatures for our stars,
clearly violating the excitation equilibrium criterium.

We also not that the $\alpha$-elements abundances seem to be on average
slightly higher than that of the iron group, in good agreement with
what is observed at these metallicity in disk stars of similar
metallicities (cf. for example Edvardsson et al. \cite{edvardsson etal
93}).

\begin{table}[hbt]
\caption{Abundances for the three Be~21 giants} \label{T-abundances}
\begin{tabular}{l@{}c@{~}c@{~}c@{~~}c@{~}c@{~}c@{~~}c@{~}c@{~}c}
\noalign{\smallskip}
\hline
\noalign{\smallskip}
Star & \multispan3 T33 & \multispan3 T27 &\multispan3 T26 \\
\Teff,\logg&\multispan3 \hfill 4600, 2.0 \hfill&
\multispan3 \hfill 4600, 2.0\hfill&
\multispan3 \hfill 4500, 2.0\hfill \\
\noalign{\smallskip}
& [X/H] & $\sigma$ & n & [X/H] & $\sigma$ & n & [X/H] & $\sigma$ & n\\
\noalign{\smallskip}
\hline
\hline
\noalign{\smallskip}
 \NLi&3.0 &--& 2& $\leq$0.5& & & $\leq$0.5 &   &  \\
\noalign{\smallskip}
\hline
\noalign{\smallskip}
  Na  I  & $-$0.33 & -- & ~1 & 0.06 & -- & ~1&+0.02 & --  &  ~1\\
  Al  I  & $-$0.38 & -- & ~1 &      &      &  &      &     &   \\
\hline
\noalign{\smallskip}
  Si  I  & $-$0.07 & --   & ~1 &$-$0.09 & --   & ~1&$-$0.47 & --   &  ~1 \\
  Ca  I  & $-$0.61 &~0.27 & ~6 &$-$0.56 &~0.25 & ~5&$-$0.73 &~0.18 &  ~7 \\
   V  I  & $-$0.59 &~0.09 & ~5 &$-$0.29 &~0.18 & ~5&$-$0.15 &~0.08 &  ~4 \\
  Ti  I  & $-$0.55 &~0.25 & ~7 &$-$0.37 &~0.26 & ~5&$-$0.29 &~0.23 &  ~7 \\
  Ti II	 &   +0.08 &~0.21 & ~2 &$-$0.33 & --   & ~1&      &      &    \\
$[\alpha$/H$]$& $-$0.56&~0.24&~19& $-$0.39&~0.25 &~16 & $-$0.42&~0.30&~20\\
\hline
\noalign{\smallskip}
  Cr  I  & $-$0.86 & --   & ~1 &$-$0.76 & --   & ~1  &      &      &    \\
  Fe  I  & $-$0.58 &~0.21 &~31 &$-$0.55 &~0.23 &~24&$-$0.44 &~0.24 &~27 \\
  Fe II  &   +0.21 &~0.08 & ~3 &  +0.07 &~0.08 & ~2&$-$0.34 &~0.26 &  ~4 \\
  Ni  I  & $-$0.56 &~0.16 & ~7 &$-$0.64 &~0.21 & ~6&$-$0.48 &~0.22 &  ~6 \\
$[$``Fe''/H$]$& $-$0.58&~0.20&~39& $-$0.57&~0.22&~31& $-$0.44&~0.24&~33\\
\noalign{\smallskip}
\hline
\noalign{\smallskip}
\end{tabular}
{\bf Note :} In addition to the individual elements abundances, we
display also for each star the mean of two groups of elements: the
$\alpha$-elements [$\alpha$/H], including Si, Ca, V and Ti, 
and the iron-group elements [``Fe''/H], including Cr, Fe, and
Ni.
\end{table}

	\subsection{Lithium} \label{lithium}

\begin{figure}[tbp]
\resizebox{\hsize}{!}{\rotatebox{-90}{\includegraphics{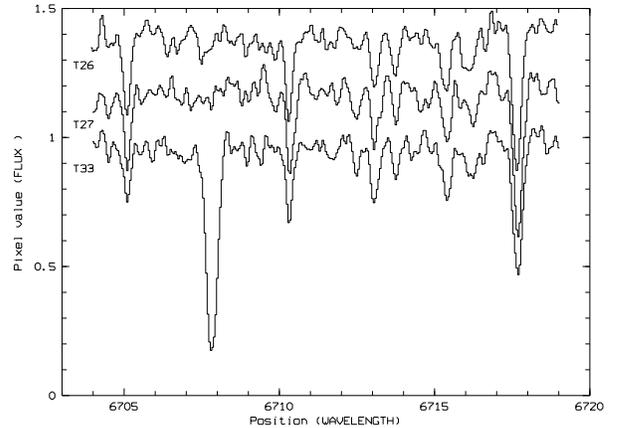}}}
\caption{Observed Lithium 6707\AA~ line in the three Be~21 giants.}
         \label{F-Li67}
   \end{figure}

\begin{figure}[tbp]
\resizebox{\hsize}{!}{\rotatebox{-90}{\includegraphics{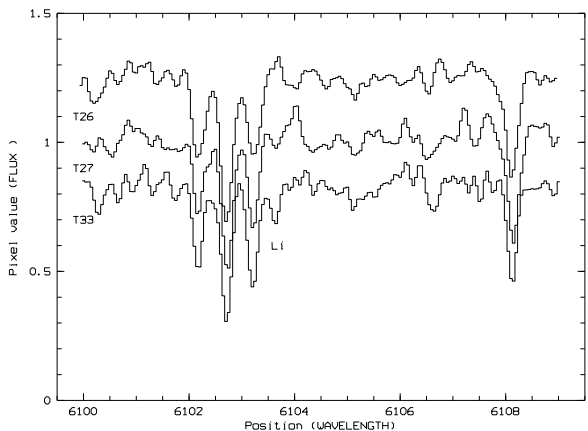}}}
\caption{Observed Lithium 6104\AA~ line in the three Be~21 giants.}
         \label{F-Li61}
   \end{figure}

The spectral region around the 6708\AA~ Li I resonance line is given in
Figure~\ref{F-Li67} for the three giants,
For the star T33 the LTE analysis gives $\NLi = 3$~dex; we caution that for
these saturated line the abundance can be much larger
than what given, in particular if an NLTE treatment is done (de la Reza
\& da Silva \cite{de la resa da silva 95}) . In any case, this implies that 
the Li abundance of this star
is comparable to the initial Li abundance of the cluster
or even higher. Since the secondary Li I line
at 6104\AA~ is also included in our spectrum, 
we checked for this line, as given in Figure~\ref{F-Li61}.
Given the limited S/N ratio of the data, all we can say is that the line is
present and the equivalent width ($W \sim 48$m\AA) is consistent with 
the abundance found from the 6708\AA~ resonance line.

For T26 and T27, only an upper limit of $\NLi = 0.5$~dex can be derived,
assuming 30m\AA~ for the detection limit of the 6708\AA~ line.


\section{Discussion and Conclusions}

The mechanisms proposed to explain SLIR giants can be basically
divided in two type: pollution and self enrichment.

Pollution can happen either via the mass transfer in a binary system
where the more massive star evolves first to a white dwarf and
pollutes its companion by novae-explosions,
or via the engulfing of a low mass companion (i.e. brown dwarf) rich in Li.
We cannot at the moment exclude any of these mechanism. 
The two spectra of T33 were too close in time to detect radial
velocity variations, and anyhow, major radial velocity variations 
are not expected when the companion is a white dwarf. 
Also, we have no access to the Be region to test
the abundance of this material, which could rule out the
BD engulfing scenario, as shown for two Lithium Rich field giants by
Castilho et al. (\cite{castilho etal 99}). But the similarity of the radial velocity of this
giant to the three others reduces the probability that it is a binary
system. 
Concerning the BD scenario, however, we note also that
the difference in Li abundance between T33 and the other two giants
is so large (a factor $\sim$1000) that the engulfing hypothesis 
would have problems in providing such an amount of Li (Siess \& Livio
\cite{siess livio 99}). Also, in such a scenario, the engulfed low-mass companion would
transfer angular momentum to the giant, leading to a substantial
spin-up of the giant. This is not observed the case of T33, which does
not seem to be rotating any faster than the two similar giants T26 and T27.

A variation of the Fowler-Cameron  self enrichment mechanism has been
proposed by Sackmann \& Boothroyd (\cite{sackmann boothroyd 99}) and it has been
advocated by Kraft et al. (\cite{kraft etal 99}) as a possible explanation of
their SLIR giant in M3.
At the moment this is an hypothesis, which
possibly also needs of some link with the presence of dust  shell
as observed  by de la Reza et al. (\cite{de la
resa etal 96}).
With respect to Kraft's SLIR giant in M3 we note that although T33
is more luminous than the cluster clump, its absolute magnitude is
about 2 magnitudes fainter than  the M3 star. 
The star in M3 is also substantially cooler.
Another comparison can be done with the other SLIR field star HD39853
(Gratton and D'Antona \cite{gratton dantona 89}): with  a metallicity similar to  T33
its Hipparcos parallax indicates that this star
is about 1 magnitude brighter and  substantially cooler than T33:
HD39853 should therefore be very similar to the Be~21 star T9,
which does not show a strong Li abundance.

A possible argument in favour of pollution could be proposed:
observations of Li in giants of clusters are limited to some tens of
objects, and, if the few cluster SLIR giants discovered so far were 
indicative of the trend, the percentage of SLIR giants in clusters 
could be higher than what is observed in the field (cf. Brown et
al. \cite{brown etal 89}). This suggests therefore that
environment may have an important role. On the other hand,
we have also to note that these clusters are typically more metal
poor than the field stars, and, for instance, the
Sackmann \& Boothroyd mechanism should work better
at lower metallicities.
However, a proper statistical analysis could only be performed if
complete samples of cluster giants were observed, which is not the case
at present, where only scarce lithium data are available for 
open cluster giants.

\begin{acknowledgements}
We thank G. Marconi for providing us with the Be~21 photometry. We
also thank R. de la Reza for his useful comments as a referee of 
the paper.
\end{acknowledgements}


\begin{thebibliography}{}
	\bibitem[1996]{alonso etal 96}
Alonso A., Aribas S., Mart\`{\i}nez-Roger C., 1996, A\&A 373, 890
	\bibitem[1998]{bessell etal 98}
Bessell M., Castelli F., Plez B., 1998, A\&A 333, 231
	\bibitem[1989]{brown etal 89}
Brown J.A., Sneden C., Lambert D.L., Dutchover E.Jr., 1989. ApJS, 71, 293
	\bibitem[1998]{carney etal 98}
Carney B., Fry A., Gonzales G., 1998, AJ 116, 2984
	\bibitem[1971]{cameron fowler 71}
Cameron A.G., Fowler W.A., 1971, ApJ 164, 111
	\bibitem[1999]{castilho etal 99}
Castilho B.V., Spite F., Barbuy B., Spite M., de Medeiros J.R.,
Gregorio-Hetem J., 1999, A\&A 345, 249
        \bibitem[1995]{de la resa da silva 95}
de la Reza R., da Silva L., 1995, ApJ 439, 917  
        \bibitem[1996]{de la resa etal 96}
de la Reza R., Drake N. A., da Silva L., 1996, ApJ 456, L115 
        \bibitem[1997]{de la resa etal 97}
de la Reza R., Drake N. A., da Silva L., Torres C. A., Martin E.L.,
1997, ApJ 482, L77	
	\bibitem[1993]{edvardsson etal 93}
Edvardsson, B., Andersen, J., Gustafsson, B., Lambert, D.L., 
Nissen, P.E., Tomkin, J., 1993, A\&A 275, 101
         \bibitem[1993]{friel janes 93}
Friel E. D., Janes K. A., 1993, A\&A 267, 75
         \bibitem[1995]{friel 95}
Friel E. D., 1995, Ann. Rev. of A\&A 33, 381
         \bibitem[1989]{gratton dantona 89}
Gratton, R.G., D'Antona, F. 1989, A\&A 215, 66
	\bibitem[1999]{kaufer etal 99}
Kaufer A., Stahl S., Tubbesing S., N\o rregaard P., Avila, G.,
Fran\c{c}ois P., Pasquini L., Pizzella A., 1999, 
The ESO Messenger 95, 8
	\bibitem[1999]{kraft etal 99}
Kraft R. P., Peterson R. C., Guhathakurta P., Sneden C., Fulbright J.,
Langer G. E., 1999, ApJL (in Press)
	\bibitem[1992]{plez etal 92}
Plez B., Brett J.M., Nordlund A., 1992, A\&A 256, 551
	\bibitem[1997]{plez 97}
Plez B., 1995 and 1997 private communication
	\bibitem[1999]{sackmann boothroyd 99}
Sackmann, I-J, Boothroyd, A.I. 1999, ApJ 510, 217
	\bibitem[1999]{siess livio 99}       
Siess L., Livio M., 1999, MNRAS 304, 925
	\bibitem[1998]{tosi etal 98}        
Tosi M., Pulone L., Marconi G., Bragaglia A., 1998, MNRAS 299, 834

\end{thebibliography}
\end{document}